\documentclass{sig-alternate}

\usepackage{etoolbox}
\makeatletter
\patchcmd{\maketitle}{\@copyrightspace}{}{}{}
\makeatother

\usepackage{balance}  
\usepackage{graphics} 
\usepackage{times}    
\usepackage{url}      
\usepackage{subfigure}
\usepackage{xspace}

\makeatletter
\def\url@leostyle{%
  \@ifundefined{selectfont}{\def\UrlFont{\sf}}{\def\UrlFont{\small\bf\ttfamily}}}
\makeatother
\def\pprw{8.5in}
\def\pprh{11in}

\usepackage[pdftex]{hyperref}
\hypersetup{
pdftitle={On the Effect of Human-Computer Interfaces on Language Expression},
pdfauthor={LaTeX},
pdfkeywords={Keyboard entry; Mobile Devices; Touch Screen; Expression; Language},
bookmarksnumbered,
pdfstartview={FitH},
colorlinks,
citecolor=black,
filecolor=black,
linkcolor=black,
urlcolor=black,
breaklinks=true,
}

\newcommand{\YA}{\mbox{Yahoo!\ Answers}\xspace}

\begin{document}

\title{On the Effect of Human-Computer Interfaces\\ on Language Expression}

\numberofauthors{3}
\author{
  \alignauthor Dan Pelleg\\
    \affaddr{Yahoo! Research}\\
		\affaddr{Haifa, Israel}\\
    \email{dpelleg@yahoo-inc.com}
  \alignauthor Elad Yom-Tov\\
    \affaddr{Microsoft Research}\\
    \affaddr{Herzeliya, Israel}\\
    \email{eladyt@microsoft.com}
  \alignauthor Evgeniy Gabrilovich\\
    \affaddr{Yahoo! Research}\\
		\affaddr{Santa Clara, CA, USA}\\
		\email{gabr@acm.org}\titlenote{This work has been done while the author was with Yahoo Research. The author's current affiliation is Google, Mountain View, CA, USA.}
}


\maketitle

\begin{abstract}
Language expression is known to be dependent on attributes intrinsic to the author. To date, however, little attention has been devoted to the effect of \emph{interfaces} used to articulate language on its expression. Here we study a large corpus of text written using different input devices and show that writers unconsciously prefer different letters depending on the interplay between their individual traits (e.g., hand laterality and injuries) and the layout of keyboards. Our results show, for the first time, how the interplay between technology and its users modifies language expression.
\end{abstract}

\section{Introduction}

Language expression is known to be dependent on attributes intrinsic to the author, such as education, gender, and age~\cite{LeaperS2004}. However, after ideas are initially formed, they are often verbalized via interaction with mechanical input devices. In this paper we show that the technology --- and, in particular, keyboard --- has a measurable influence on language expression. Specifically, we show that writers\footnote{We use the term "writer" in the broad sense to refer to people who write texts, not solely to authors of manuscripts of substantial size or scope.} unconsciously prefer different letters depending on the interplay between their individual traits and the layout of the keyboard. We also show that the same writer may have different ``personas" (as evidenced by different letter preferences) when interacting with different kinds of keyboards.

Arguably, the most common text input mechanism today is the QWERTY keyboard. It was designed to distribute frequently-used letter pairs between the two hands in order to increase typing speed. The keyboard is asymmetrical, with more letters located to the left of the midline~\cite{JasminC2012}. Recently, it has been shown that letter location on the keyboard is associated with sentiment, and words containing letters from the right side of the keyboard are estimated as having more positive meaning~\cite{JasminC2012}. However, to the best of our knowledge, it has not been previously known whether writers actually modify their language expression when interacting with different keyboards.

We examined publicly accessible free-form text passages written voluntarily
at \YA, a social question-answering web site. The passages were typed using
various input devices: desktop keyboards\footnote{When referring to desktop
  computers, we also include laptops since both have full-size mechanical
  keyboards, and also we have no way to distinguish between them in the server
  logs.}, smartphones, and tablet computers. Observe that when users
interact with a full-size keyboard on desktop computers, they usually type
with both hands. However, this practice changes when users interact with
tablet computers or smartphones. In these cases, users often hold the
device with one hand and type with the other hand (often using a single
finger), or hold the device with both hands and type with the thumbs. Our
findings indicate that when the same users operate different devices, they
exhibit statistically significant preferences for different letters on each
device. Importantly, this observation remains valid even when we focus on
sub-populations of users, namely, those who have voluntarily
self-identified to be left- or right-handed, or those who have
self-declared to be suffering from Carpal Tunnel Syndrome (CTS). These
special populations are interesting because they have notably different
patterns of keyboard use compared to the general population. Nonetheless,
we clearly see that users in these groups also exhibit statistically
significant preferences for different letters on different devices.

Our findings indicate that human-computer interfaces, and specifically
keyboards, are not language-neutral, and instead have measurable effect on
language expression by their users. To the best of our knowledge, this
effect has not been discovered in prior studies. Further research is
required to quantify the differences in the perception of language produced on
different input devices. We believe future research should also focus on
developing language-neutral input interfaces, which do not noticeably
affect language expression.


\bigskip
\bigskip

\section{Methodology}

We report the results of three experiments comparing different populations active on \YA. Previous research~\cite{SocialCom2012:Pelleg} showed that users who state their preferences and physical attributes on \YA are overwhelmingly accurate in the data they provide. Therefore, in this work we assumed that users' statements about their personal traits, such as hand laterality or being affected by Carpal Tunnel Syndrome, are accurate.

In the first experiment, we analyzed English-language posts entered by \YA users in 2012. There were about 1.8~million such posts, and Table~\ref{tab:stats} describes their properties,
broken by input device. In general, the numbers observed for tablet devices in Table~\ref{tab:stats} are
between those for desktop and mobile phones, therefore, in the discussion below we mainly compare desktops and mobile phones. Posts from desktop
computers are typically longer by 67 characters or 11 words, when compared
to posts from mobile phones. This is not surprising when one compares the convenience of the
full-size keyboard of desktop computers with the limitations of small keyboards available on handheld devices
(either on-screen or mechanical). The latter usually require more keystrokes to type punctuation and other special characters, and only allows typing using one or at most two fingers at once. We also see that the number of URLs posted from mobile phones is much
lower, which might be due to the difficulty of switching to a different
application, such as a web browser or mail client, in order to retrieve the URL. Similarly, the frequency of commas is reduced by 50\% on mobile
devices, and we note that these require one to switch away from the
standard keyboard layout, at least on the popular iPhone
devices\footnote{On Android devices, both comma and period are on the main keyboard layout screen.}. The frequency of periods is also reduced, but
only by 33\% - perhaps due to the autocomplete feature (replacing a double
space with a period). Surprisingly, emoticons are more frequent on mobile
phones compared to desktop computers. We believe that in this case the need for expressing emotion in casual conversations is stronger than the
added difficulty of typing less accessible characters.

However, such analysis ignores several confounding variables. For example, the ownership rates for web-capable phones and corresponding data plans might vary by age and level of education. On top of that, the propensity of a user to actually browse the web from their mobile device, given that they have a suitable device, might depend on age or gender. These kinds of biases could easily lead to differences in the texts typed on various devices, which cannot be solely attributed to the nature of interaction with the particular device.

To overcome this difficulty, we identified 479 U.S.-based users who contributed text from a desktop computer and, at distinct occasions, also from a mobile phone or a tablet computer. We can now perform paired tests of contributions from these users, eliminating the confounding variables such as age, gender, or level of income. We extracted posts from these users, and removed the URLs from them\footnote{We excluded URLs because users effectively have no choice of letters when typing them.}. Then, we computed the probability of using each letter for each device. We used a paired sign test at the significance level of 0.05 with Dunn-Sidak correction~\cite{Abdi2007} to evaluate which letters were used at significantly different levels on each mobile device (smartphone or tablet), compared to the desktop.

In our second experiment, we looked at \YA posts from users who self-identified as suffering from Carpal Tunnel Syndrome. CTS is a medical condition causing wrist and hand pain and difficulty in typing. A typical user would disclose this fact when posting a relevant question in the ``injuries'' or ``pain \& pain management'' categories. Using text matching on either the disease acronym or its full name, we identified 2636 such users. For these users we extracted all the questions and answers they contributed to \YA over a span of 6 years (2007---2012). As a baseline, we used the text from a random sample comprised of the same number of users.

In the third experiment, we extracted text contributed by users who voluntarily identified their handedness (1081 left-handed and 1555 right-handed users). As an example, users often disclose their handedness when posting a question about a particular sport technique, e.g., serving a ball in tennis. To this end, we extracted the text from all questions and answers posted by the users who explicitly stated that they were right- or left-handed, over a span of 6 years (2007---2012).

We associated each letter with its Euclidian distance from the center of a standard English-language QWERTY keyboard. We hypothesized that the dominant hand may be placed at different locations on the keyboard depending on hand laterality, making different areas of the keyboard more accessible. Therefore, we tested each possible location on the keyboard as a potential center. For each such location we measured the average distance (from the keyboard center) of the keys used by left-handed users, divided by that of right-handed users. We assessed the statistical significance of the difference using the Kruskal-Wallis test.

\section{Results}

\begin{figure*}[t]
    \centering
    \subfigure[Mobile phone vs. desktop]
    {
        \includegraphics[width=1.0\textwidth]{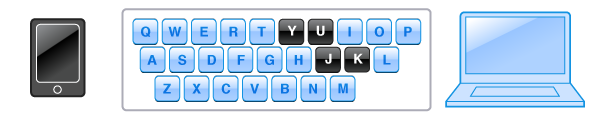}
        \label{fig:mobile-vs-pc}
    }
    \subfigure[Tablet computer vs. desktop]
    {
        \includegraphics[width=1.0\textwidth]{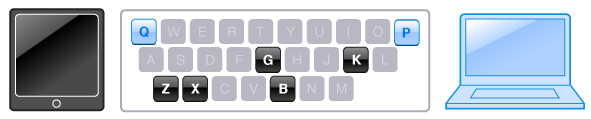}
        \label{fig:tablet-vs-pc}
    }
\caption{Letter preferences on different pairs of devices. Letters shown on the blue background are preferred by desktop users, while letters shown on the black background are preferred by mobile phone / tablet users. In diagram (b), letters shown in light gray background are not preferred by either population.}
\label{fig:keyboard_usage}
\end{figure*}

Our first experiment was designed to measure the effect of different input
devices on the same person. Here we analyzed, for each writer, the text she
entered using a desktop computer and either a mobile phone or a tablet
computer. We found statistically significant differences in the usage of
letters, as follows. On mobile phones, the letters KYUJ were more
frequently used, compared to their use by the same writers on a desktop
computer. Conversely, the letters (statistically significantly) more often used
on desktop computers as compared to mobile were
QIBSVXLWENOGTHPRADCFMZ. Comparing tablet computers to desktops, we found
that the letters XBGKZ were more frequently used on tablets and the letters QP were more
frequently on desktop computers. See Figure~\ref{fig:keyboard_usage} for an
illustration.

Thus, within-subject preference for letters on mobile phones is for letters on the right-hand side of the keyboard (KYUJ), which we attribute to the common practice of typing with the thumb of the same hand (usually right) that is holding the mobile phone, or alternatively, holding the device with the left hand and typing with the right index finger. In tablets, 3 out of 5 preferred letters were in the bottom row of the keyboard, while the least preferred letters were all in the top row. We believe this is explained by the common typing position where one hand performs most of the typing while hovering above the screen surface, which makes extending the arm towards the top row more difficult.

In the second experiment, we examined inter-subject variability in text contributed by writers who voluntarily stated that they were affected by the Carpal Tunnel Syndrome. When compared to the control population, these writers preferred keys farther from the center of the keyboard, when the latter is defined\footnote{See the last paragraph of the previous section for an explanation why multiple alternative positions of the center of the keyboard had to be considered.} at seven letters from the left hand side of the keyboard ($p=0.0013$, Kruskal-Wallis test). This preference is greater by 0.7\% in the CTS-affected population, compared to the control. These findings indicate that the layout of the keyboard influences letter selection for people afflicted by CTS, and may be due to CTS sufferers' using only one hand to type.

In the final experiment, we examined inter-subject variability in text
expression by people of different hand laterality. Using the same
methodology as above, we discovered that left-handed writers tend to prefer
keys farther from the center of the keyboard, when the latter is defined at
six letters from the left-hand side of the keyboard, compared to right
handed writers ($p=0.03$, Kruskal-Wallis test). This preference is greater
by 0.3\% in left-handed writers, compared to right-handed writers.

We also tallied the occurrences of consecutive pairs of letters,
where each could be on the left (L) or right (R) side of the
keyboard\footnote{The space bar was considered to be on neither side.}. For
each post, we normalized the counts by the total number of pairs ($N-1$ pairs for
a post of length $N$). Our findings are summarized in Table~\ref{tab:pairsides}. We found that for RR
pairs (two consecutive letters on the right side of the keyboard), the left-handed group
types them more often than the right-handed group. For LL pairs, the
right-handed group types them more often than the left-handed group. For both LR
and RL pairs, the right-handed group types them more often than the left-handed
group. These results could be summarized by the following rules:
\begin{enumerate}
\item For letter pairs on the same side of the keyboard, the group that
  types them more often is the one with the dominant hand on the opposite side.
\item For letter pairs on opposing sides of the keyboard, right-handed
  people type them more often than left-handed people.
\end{enumerate}

\begin{table}[t]
\centering
\begin{tabular}{l|c|c|c}
\bf{Metric} & \bf{Desktop} &  \bf{Mobile} &  \bf{Tablet} \\
\hline
length in characters & 215   & 148          &  165        \\
length in words      & 39    & 28           &  30         \\
characters per word & 5.7  & 5.3          & 5.3         \\
URL count          & 0.07  & 0.013        & 0.05        \\
emoticons per word & 0.079 & 0.084        & 0.075       \\
period count          & 3.3   & 2.2          & 2.5         \\
comma count        & 1.6   & 0.8          & 1.0         \\
\end{tabular}
\caption{Statistics on \YA posts for U.S.\ users, broken down by device. In each row, the difference is statistically significant at the level $p<0.001$ for all three pairs.}
\label{tab:stats}
\end{table}

\begin{table}[t]
\centering
\begin{tabular}{l|c|c|c|c}
 & \multicolumn{3}{c}{Type of key pair} \\
Population &  LL & LR & RL & RR \\
\hline
Control &      16.65\% &  16.34\% & 16.07\% &  15.44\% \\
Left-handed &  16.44\% &  16.32\% & 15.86\% &  15.70\% \\
Right-handed & 16.51\% &  16.34\% & 15.90\% &  15.53\% \\
\end{tabular}
\caption{Frequencies for typing consecutive pairs of keys from the left (L)
  and right (R) sides of the keyboard. Differences are statistically
  significant at the level $p<10^{-10}$, with the exception of the LR
  difference between the control and right-handed populations.}
\label{tab:pairsides}
\end{table}


\section{Conclusion}

From typewriters to desktop computers and then to laptops, textual input has been dominated by mechanical QWERTY keyboards. Today, however, touch interfaces and on-screen keyboards are becoming more commonplace. On-screen interfaces are no longer limited to tablet computers or mobile phones, and nowadays appear more often in devices that previously lacked rich input modes, such as navigation systems, ticket kiosks, and ``smart'' TVs. Consequently, we believe the time is ripe to revisit some of the latent assumptions about text input, in order to foster better design of these new interfaces.

In this work, we studied how users input text when interacting with different devices. Specifically, we analyzed users' postings to \YA made from desktop computers, tablets, and mobile phones. Our findings indicate, for the first time, that \emph{text input devices modify language expression in a measurable way}. This observation remains valid when the same person uses different devices, for right- and left-handed writers, and for healthy individuals compared to those affected by CTS. The latter populations of users are particularly notable, because they have developed their own, distinctive ways of using the keyboard, as influenced by their individual traits. Nonetheless, despite the variety in their physical traits, all examined user populations have one thing in common, as they clearly adapt to different input modalities, and exhibit statistically significant preferences for different letters on different devices.

We envision two promising avenues for future research on the topic. First, it is interesting to understand whether the differences in language expression that we have identified actually lead to different perception by the readers of the texts written on different input devices. Second, it remains an open question whether it is possible to design unbiased devices for mechanical text entry, which will not affect language expression for various user populations.


%
%
%
%
%

\nocite{NYT2007:Pogue}

\bibliographystyle{abbrv}
\bibliography{handedness}
\end{document}